\def \pd{\partial}
\def\Jhat{{\hat J}}
\def \th{\theta}
\def \ph{\phi}

\def \Ph{\Phi}
\font \bigbf=cmbx10 scaled \magstep1
\tolerance=500
\magnification=1200
{\nopagenumbers
\line{\hfil}
\line {\hfil LTH 281}
\line {\hfil March 1992}
\line {\hfil Revised Version July 2002}
\line {\hfil hep-ph/0111190}
\vskip .5in
\line{\hfil \bigbf THE STANDARD MODEL EFFECTIVE\hfil}
\line{\hfil \bigbf POTENTIAL AT TWO LOOPS\hfil}
\vskip 1in
\line{\hfil {\bf C. Ford}
\footnote{*}{ Current address:
Instituut Lorentz for Theoretical Physics, Niels Bohrweg 2, 
NL-2300  RA Leiden, The Netherlands}
{\bf and I. Jack\hfil}}
\line{\hfil}
\line{\hfil \it Dept. of Mathematical Sciences, 
University of Liverpool, Liverpool L69 3BX,
UK\hfil}
\vskip .5in
\line{\hfil\bf D. R. T. Jones\hfil}
\line {\hfil}
\line{\hfil \it ITP, University of California at Santa Barbara,
CA 93106, USA
\footnote{**}{Until 28th May 1992.}\hfil}
\line {\hfil}
\line{\hfil\it Dept. of Mathematical Sciences, 
University of Liverpool, Liverpool L69 3BX, UK
\footnote{***}{Permanent address.}\hfil}
\vskip 1in
\line{\hfil\bf Abstract\hfil}
 \line{\hfil}
We calculate the standard model effective potential to two loops using
minimal subtraction, and use the result to deduce the two-loop
beta-functions for the scalar $m^2$ and quartic self-interaction.
 
\vfil\eject}
\pageno=1
\line{\bigbf 1. Introduction \hfil}
\line{\hfil}
The role of the effective potential $V(\ph)$ in determining the
nature of the vacuum in renormalisable field theories was emphasized
in the classic paper of Coleman and Weinberg[1] (CW). Their
particular interest was in the special case when the renormalised
value of the $(\hbox{mass})^2$ parameter, $m^2$, of the scalar fields
was zero. They were able to demonstrate the occurrence of
spontaneous symmetry breaking through radiative corrections for this
case, as long as gauge fields are present. This scenario however, is
excluded by current experimental limits on $m_H$ (the Higgs mass)
and $m_t$ (the top mass).
 
In the $m^2 <0$ case, radiative corrections are still important
in determining whether the tree minimum of  $V$ corresponds to the
true ground state of the theory. ( For a review and references, see
ref.~[2] ). The one-loop Yukawa coupling contribution to $V$ tends to
destabilise the vacuum, and consequently leads to an upper bound for
             $m_t$ as a function of $m_H$ [3].
 
The analysis described in ref.~[3] was based on a
renormalisation-group (RG) \lq\lq improved" form [1,4] of $V$ including
one-loop corrections. Here        we present the results of the
two-loop radiative corrections to $V$, $V^{(2)}$, with a view to
possible refinement of the bounds described in ref.~[3].
 
Another motivation for our calculation is as follows: Given the
standard model 2-loop $\beta$-function, the dependence of
$V^{(2)}$ on the renormalisation scale $\mu$ may be readily
inferred from the fact that $V$ satisfies a RG equation. (This
calculation was in fact attempted in ref.~[5]; we will comment
later on this paper). We shall instead calculate the full
$V^{(2)}$, using minimal subtraction ($MS$) throughout, and use the
results to infer the 2-loop $\beta$-functions for $m^2$ ($\beta^{(2)}_{m^2}$)
  and the quartic Higgs coupling $\lambda$ ($\beta^{(2)}_\lambda$).   As we shall
see, this procedure will expose some minor errors in the expression for
the SM $\beta^{(2)}_\lambda   $ given in ref.~[6]. As for $\beta^{(2)}_{m^2}$
, as far as we know it has not previously appeared in the literature.
As already mentioned in ref.~[7], it is essential that $V^{(1)}$ be
calculated using $MS$  in order that $V^{(2)}$ be consistent with the
RG equation with the $MS$ $\beta$-functions;   the  $\beta$-functions  in a
multi-coupling constant theory are scheme dependent at the two-loop
level [8]. Thus although there exists [9] (in the special case
($h=m^2=0$)) a calculation of $V^{(2)}$,  this cannot be used for
comparison with the standard model $\beta^{(2)}_{\lambda }$, since $MS$
was not employed.
 
The plan of this paper is as follows. In section 2 we describe briefly
the effective potential formalism, give the result for the one-loop
correction, $V^{(1)}$, and introduce some notation. In section 3 we
explain our procedure for finding $V^{(2)}$ within the $MS$ scheme,
and in section 4 we apply a novel method to the evaluation of the
pertinent Feynman integral. The method involves the use of
differential equations [10] with the added refinement of the method of
characteristics. Section 5 consists of a summary of the results, while
in section 6 we show how the fact that $V(\ph)$ satisfies a RG
equation leads to a determination of $\beta^{(2)}_{m^2}$ and
$\beta^{(2)}_{\lambda}$. Finally in section 7 we discuss applications.
\vskip 1in
\line {\bigbf 2. The one-loop effective potential.\hfil}
\line{\hfil}
The effective potential formalism of CW and the functional
refinements introduced by Jackiw[11] are too well known to require any
but the briefest review here. In general one shifts scalar fields as
follows:
$$\ph(x) \rightarrow \ph+\ph_q(x) \eqno (2.1)$$
where $\ph$ is $x$-independent.Then the effective potential $V(\ph)$ is
given by the sum of \it vacuum \/\rm graphs, with $\ph$-dependent
propagators. Equivalently, one can calculate graphs with a single
$\ph_q$ external field, which, it is easy to show, leads to a
determination of $\pd V     /\pd\ph$. The latter method is perhaps
simpler at one loop, but not beyond since it leads to more Feynman
diagrams.(Here we differ as a matter of opinion from ref.~[9]).
 
In the SM case, one can exploit gauge invariance to perform the shift
of Eq.~(2.1) on one only of the four scalar fields:
$$\Ph(x)\rightarrow\pmatrix{0\cr \ph\cr}+\pmatrix{G^{\pm}(x)\cr
{1\over{\sqrt2}}(H(x)+iG(x))\cr}.  \eqno (2.2)$$
We must also choose a gauge; the
'tHooft-Landau gauge is the most convenient one.(In fact $V(\ph)$ is
gauge invariant only at its extrema; this gauge is a good one in the
sense of ref.~[12]). In this gauge the $W$, $Z$ and $\gamma$ propagators
are transverse, and the associated ghosts are massless and couple only
to the gauge fields; the \lq\lq would be Goldstone" bosons $G^{\pm}$,$G$
have a common mass deriving from the scalar potential only.
 
At the tree level the effective potential is $V^{(0)}(\ph)$, given by
$$V^{(0)}(\ph)={m^2\over2}\ph^2+{\lambda\over{24}}\ph^4 \eqno(2.3)$$
(Note that our definition of $\lambda$ differs by a factor of 3 from that
of ref.~[6].)
 
The result of the one loop  calculation is:
$$\eqalignno{
\kappa V^{(1)} & ={H^2\over4}\bigl(\overline{\ln}H-{3\over2}\bigr)+{3G^2
\over4}\bigl( \overline{\ln}G-{3\over2}\bigr)-3T^2\bigl(\overline{\ln}T-
{3\over2}\bigr)\cr
&+{3W^2\over2}\bigl(\overline{\ln}W-{5\over6}\bigr)
+{3Z^2\over4}\bigl(\overline{\ln}Z -{5\over6}\bigr)&(2.4)\cr}$$
where
$$\kappa=16\pi^2\hbox{, }H=m^2+{\lambda\over2}\ph^2\hbox{, }T={h^2\over2}
\ph^2$$
$$G=m^2+{\lambda\over6}\ph^2\hbox{, }W={g^2\over4}\ph^2\hbox{, }
Z={(g^2+g'^2)\over4}\ph^2$$
$$\hbox{and }\overline{\ln}X=\ln{X\over{\mu^2}}+\gamma-\ln4\pi,$$
$\gamma$
being Euler's constant. (Note that the sign of the
$\ln4\pi$ term is rendered
incorrectly in ref.~[7], Eq.~(4)).
    Here
     $h$ is the top quark Yukawa coupling (we neglect other Yukawa
couplings throughout).
 
If $m^2<0$, then at the minima of $V^{(0)}(\ph)$ we have $G=0$ and
$H$, $T$, $W$, $Z$ become the tree level $(masses)^2$ of the Higgs,
top quark, W and Z bosons respectively. The non logarithmic terms in
Eq.~(2.4) may be altered (or indeed removed) by a change in
subtraction scheme. Since, however, we wish to consider the RG
equation with the $MS$ scheme, it is essential that we retain them. (
The $\overline{MS}$ scheme corresponds, of course, to simply replacing
$\overline{\ln}X$ by $\ln( X/\mu^2)$ throughout; this would not affect
the RG analysis since all RG functions are identical in $MS$ and
$\overline{MS}$[13]).
\vskip 1in
\line {\bigbf 3. The two-loop calculation: preliminaries.\hfil}
\line{\hfil}
In this section we outline our basic strategy for the calculation.
By elementary manipulation we can reduce each individual Feynman
diagram to a sum of integrals of the form of either:
$$    I(x,y,z)={(\mu^2)^{2\epsilon}\over{(2\pi)^d}}\int{d^dk\,d^dq\over
{(k^2+x)(q^2+y)((k+q)^2+z)}}\eqno(3.1)$$
or
$$J(x,y)=J(x)J(y)\eqno(3.2)$$
where
$$       J(x)={(\mu^2)^{\epsilon}\over{(2\pi)^d}}
\int{d^dk\over{k^2+x}}={(\mu^2)^{\epsilon}\over
{(4\pi)^{d\over2}}}\Gamma(1-{d\over2})x^{{d\over2}-1}\eqno(3.3)$$
and we define $d=4-2\epsilon$. (We work in Euclidean space throughout.)
The
    evaluation of $I(x,y,z)$ is non-trivial, and will be the subject of
Section 4.
 
We turn first
to the subject of renormalisation. We choose to work throughout
 with \it renormalised \/\rm  parameters $g_3$, $g$, $g'$, $m^2$, $h$,
and $\lambda$. ($g_3$, $g$, $g'$ are the three gauge couplings, with the usual
conventions). Then rather than compute separately a set of one-loop
diagrams with counter-term insertions, we subtract (minimally) the
sub-divergences \it diagram by diagram \rm. $V^{(2)}$ is then obtained
by simply taking the finite parts of the resulting expressions,
discarding the $1/\epsilon^2$, $1/\epsilon$ poles which are
automatically cancelled by the usual renormalisation constants
(which we need not calculate). For graphs not involving vector bosons,
 this procedure amounts simply to the replacement of $I(x,y,z)$ and
$J(x,y)$ by their subtracted values:
$$\eqalignno
{I(x,y,z)&\rightarrow\hat I(x,y,z)=I(x,y,z)-{1\over
{\kappa\epsilon}}\bigl(J(x)+J(y)+J(z)\big)&(3.4)\cr J(x,y)&\rightarrow\hat
J(x,y)=J(x,y)+{1\over{\kappa\epsilon}}(xJ(y)+yJ(x)).&(3.5)\cr}
$$
There is one complication, however. When vector bosons are present, the
algebra involved in reducing the graph to dependence on $I$ and $J$
may produce explicit factors of $d$. This gives rise to an apparent
ambiguity: what is the subtracted form of $dI$? The answer is that
the result depends on which subgraph produced the factor of $d$. Thus
when vector fields are present we {\it must\/} explicitly evaluate the
contribution of the subtractions to each graph. This is still,
however, much easier than explicitly considering counter-term insertions.
 Because we work in the 'tHooft-Landau gauge, the gauge parameter is
unrenormalised and so creates no difficulties.
 
We conclude this section with the results for $J$ which will be
relevant subsequently. It is straightforward to show from (3.3) that:
$$\eqalignno{\kappa^2\hat J(x,y)&=-{xy\over{\epsilon^2}}
+xy\big(1-\overline{\ln}x
-\overline{\ln}y+\overline{\ln}x\overline{\ln}y\bigr),&(3.6)\cr
 \kappa J(x)&=-{x\over{\epsilon}}+x\bigl(\overline{\ln}x-1\bigr),&(3.7)
 \cr\kappa^2\epsilon J(x,y)&={xy\over {\epsilon}}+xy\bigl(2-\overline{\ln}x-
\overline{\ln}y\bigr).&(3.8)  \cr }$$
It is the finite part of each expression that is substituted in our
expressions for $V^{(2)}$ in Section 5. Terms of the form of Eq.~(3.7)
and Eq.~(3.8) appear in connection with the $d$ dependence discussed
above.
\vskip 1in
\line{\bigbf 4. Evaluation of I(x,y,z)   \hfil}
\line{\hfil}
In this section we indicate how the differential equations method in
conjunction with the method of characteristics leads to a simple form
for the integral $I(x,y,z)$. Of course there has been much effort
expended on higher loop Feynman integrals, and many powerful
techniques have been developed (for some examples see ref.~[14]);
nevertheless we feel that our method here has some interesting features
and is worth presenting in detail.
 
We start from the identity:
$$0=\int d^dk\,d^dq\,
{\pd \over {\pd k^{\mu}}}{k^{\mu}\over{(k^2+x)(q^2+y)((q+k)^2+z)}}.
\eqno (4.1)$$
From Eq.~(4.1) it follows that
$$2x{\pd I\over {\pd x}}+(z+x-y){\pd I\over{\pd y}}=(d-3)I +K_1(x,y,z)
\eqno (4.2)$$
where
$$K_1(x,y,z)=-{\pd J(z)\over {\pd z}}\big(J(x)-J(y)\bigr).$$
Now from Eq.~(4.2) and similar equations produced by $(x,y,z)$
permutations, one can eliminate $\pd I/\pd y$ and $\pd I
/\pd z$ and then solve the resulting equation in a similar
way to that adopted in ref.~[7]. It turns out, however,  that a more
elegant solution follows if we start with the following equation:
$$(y-z){\pd I\over {\pd x}}+(z-x){\pd I\over {\pd y}}+(x-y)
{\pd I\over{\pd z}}=K(x,y,z)\eqno (4.3)$$
where
$$\eqalignno{K(x,y,z) &= K_1(x,y,z)+K_1(y,z,x)+K_1(z,x,y)\cr&\cr &= 
-\big((z-x)(zx)^{-\epsilon}
+(x-y)(xy)^{-\epsilon}+(y-z)(yz)^{-\epsilon}\big)\Gamma' &(4.4)\cr}$$
and
   $$\Gamma'=(\mu^2)^{2\epsilon}\Gamma(2-{d\over 2})\Gamma(1-{d\over2})(4\pi)^{-d}.$$
 
The method of characteristics involves solving a system of ordinary
differential equations, as follows:
$$dt={dx\over{y-z}}={dy\over{z-x}}={dz\over{x-y}}={dI\over K}\eqno (4.5)
$$
subject to initial conditions which we shall choose to be $x=X$, $y=Y$
and $z=0$ (at $t=0$), and we will suppose without loss of generality
that $X\geq Y$.We then have:
$$
I(x,y,z)=I(X,Y,0)+\int^t_0dt' \,K(x(t'),y(t'),z(t')).\eqno(4.6)
$$
Using Eq.~(4.4) and Eq.~(4.5), we can rewrite Eq.~(4.6) as
$$I(x,y,z)=I(X,Y,0)-\Gamma'
\biggl[\int^x_Xdx\,(yz)^{-\epsilon}+\int^y_Ydy\,(zx)^{-\epsilon}+
\int^z_0dz\,(xy)^{-\epsilon}\biggr].\eqno (4.7)$$
Now from Eq.~(4.5) it follows that {\it for all\/} $t$
$$
\eqalignno{x^2+y^2+z^2&=d^2=X^2+Y^2 &(4.8)\cr x+y+z&=c=X+Y &(4.9)\cr}
$$
where $c$ and $d$ are constants. Therefore
$$xy=z^2-cz+{1\over2}(c^2-d^2)\eqno (4.10)$$
with similar equations for $yz$ and $zx$. Hence Eq.~(4.7) becomes
$$I(x,y,z)=I(X,Y,0)-\Gamma'\Biggl[\biggl(\int^{x-{c\over2}}_a+\int^a_{{c\over
2}-y}+\int^{c\over2}_{{c\over2}-z}\biggr)\ ds\,(s^2-a^2)^{-\epsilon}\Biggr]
\eqno(4.11)$$
where
     $$a=\sqrt{{d^2\over 2}-{c^2\over 4}}={1\over 2}(X-Y)=
                                {1\over2}
(x^2+y^2+z^2-2xy-2yz-2zx)^{1\over2}.\eqno (4.12)$$
Now $I(X,Y,0)$ is also tricky to evaluate by elementary methods; but by
employing the method of characteristics once again, one can show that
$$I(X,Y,0)=I(X-Y,0,0)+\Gamma'\int^{{1\over2}(X+Y)}_{{1\over2}(X-Y)}ds\,
\bigl[s^2-{1\over4}(X-Y)^2\bigr]^{-\epsilon}.\eqno (4.13)$$
Substituting Eq.~(4.13) in Eq.~(4.11) and using Eq.~(4.9) and Eq.~(4.12)
we obtain
$$I(x,y,z)=I(2a,0,0)+\Gamma'\Bigl[F({c\over2}-y)
+F({c\over2}-z)-F(x-{c\over2})
\Bigr]\eqno(4.14)$$
where
$$F(w)=\int^w_ads\,(s^2-a^2)^{-\epsilon}.\eqno(4.15)$$
Since $I(2a,0,0)$ can be evaluated by elementary methods, we have
reduced the problem to a single integral, $F(w)$. In spite of
appearances, Eq.~(4.14) is symmetric with respect to $(x,y,z)$
permutations, since it is easy to show that, for example, 
$$F({c\over 2}-y)-F(x-{c\over2})=F({c\over2}-x)-F(y-{c\over2}).$$
However Eq.~(4.14) 
is only valid in the region $a^2\geq 0$. In the region $a^2\leq 0$, it
is possible to derive the following form of the solution:
$$I(x,y,z)=-I(2b,0,0)\sin{\pi d\over 2} +\Gamma'\Bigl[G({c\over2}-x)+G({c
\over2}-y)+G({c\over 2}-z)\Bigr]\eqno(4.16)$$
where
$$G(w)=\int^w_0ds\,(s^2+b^2)^{-\epsilon}\quad \hbox{and}\quad 
b^2=-a^2.\eqno(4.17)$$
It is a nice exercise to show that for $z=x$, Eq.~(4.16) can be rewritten
as Eq.~(11a) of ref.~[7]. Note that for $a^2=0$, which in $x$,$y$,$z$
space is a cone with its apex at the origin, the integral
is trivial.
 
Writing
$$(s^2+b^2)^{-\epsilon}=1-\epsilon\ln (s^2+b^2)+
{1\over 2}\epsilon^2\ln^2(s^2+b^2)+...
\eqno (4.18)$$
and using
$$(4\pi)^d
I(x,0,0)={\Gamma(2-{d\over2})
\Gamma(3-d)\Gamma({d\over2}-1)^2\over{\Gamma({d\over2})}}
\bigl({x\over{\mu^2}}\bigr) ^{d-3}\eqno(4.19)$$
we obtain from Eq.~(4.16) that
$$\eqalignno{\kappa^2I(x,y,z)= &-{c\over {2\epsilon^2}}
-{1\over{\epsilon}}\bigl(
{3c\over2}-L_1\bigl)\cr&\cr&-{1\over2}\Bigl[L_2-6L_1+(y+z-x)
\overline{\ln}y \overline{\ln}z\cr &+(z+x-y)
\overline {\ln}z\overline{\ln}x +(y+x-z)\overline{\ln}y
\overline{\ln}x \cr &+\xi (x,y,z)+c(7+\zeta(2))\Bigr]&(4.20)}$$

where
$$L_m=x\overline {\ln}^mx+y\overline{\ln}^my+z\overline{\ln}^mz\eqno(4.21)
$$
and
$$ \xi  (x,y,z)=8b\Bigl[ L(\th_x)+L(\th_y)+L(\th_z)-{\pi\over2}\ln2\Bigr].
\eqno (4.22)$$
$L(t)$ is Lobachevskiy's function[15], defined as
$$L(t)=-\int^t_0dx\,\ln\cos x.\eqno(4.23)$$
The angles $\th_x$, $\th_y$, $\th_z$ are given by
$$ \th_x=\tan^{-1}\bigl({{c\over2}-x\over b}\bigr) \hbox{ etc.}
\eqno (4.24)$$
 
Eq~(4.22) is valid only in the region $a^2\leq 0$ (ie. inside the
cone). For $a^2>0$, we obtain from Eq.~(4.14)
a result identical to Eq.~(4.20)
except that now
$$\xi  (x,y,z)=8a\Bigl[-M(-\ph_x)+M(\ph_y)+M(\ph_z)\Big]\eqno(4.25)$$
where
$$M(t)=-\int^t_0dx\,\ln\sinh x $$
and $\ph_x$, $\ph_y$, $\ph_z$ are given by
$$\ph_x=\coth^{-1}\bigl({{c\over2}-x\over a}\bigl)\quad \hbox{ etc.}$$
$\xi  (x,y,z)$ is $\mu$-independent and therefore plays no role in the
RG analysis of section 6.
 
It is interesting to compare the results Eq.~(4.20) and Eq.~(4.22)
with those of ref.~[7] for the special case $z=x$. In that case we have
$$\sin \th_x=\sin \th_z={1\over 2}\bigl({y\over x}\bigr)^{1\over2}$$
$$\sin \th_y=1-{y\over{2x}}$$
and using the identity
$$
2L\biggl(\sin^{-1}\Bigl({\sqrt t\over2}\Bigr)\biggr)+L\biggl(\sin^{-1}
\Bigl(1-{t\over2}\Bigr)\biggr)={\pi\over2}\ln 2+2\int^{\sin^{-1}\bigl({
\sqrt t\over2}\bigr)}_0 \ln (2\sin x)\, dx\eqno (4.26)
$$
it is easy to show that Eq.~(4.20) reduces to Eq.~(12) of ref.~[7].
A similar process works in the $a^2>0$ case.
 
We will make extensive use of the form $I$ takes when its subdivergences
are subtracted, $\hat I$. From Eq.~(3.4) we find that
$$\hat I=I+{1\over {\kappa^2\epsilon^2}}\bigl(c+(c-L_1)\epsilon +({1\over2}L_2-L_1+c
+{1\over2}c\zeta(2))\epsilon^2+...\bigr).\eqno(4.27)$$
It is the \it finite part \/\rm of this expression which we use
subsequently.
\vskip 1in
\line{\bigbf 5. Two-loop results.\hfil}
\line{\hfil}
In this section we present our results for the various contributions
to $V^{(2)}$ in (what we hope is) a clear and systematic manner. It
is natural to divide the calculation into parts according to the nature
of the contributing fields, thus
$$ V^{(2)}=V_S+V_{SF}+V_{SV}+V_{FV}+V_V\eqno (5.1)$$
where $S$, $F$, $V$ denote scalar, fermion and vector fields
respectively.
 
The Feynman rules of the standard model are well known; for the
calculation of the effective potential we must only remember that we
are not calculating at the tree minimum of the potential. Apart from
giving the $G$, $G^{\pm}$ bosons a non-zero mass, this makes little
difference. The results are as follows:
$$\eqalignno{V_S= &{-\lambda^2\ph^2\over 12}\bigl[ \hat I(H,H,H)+\hat I(H,G,G)
\bigr] \cr &+{\lambda\over 8}\bigl[\hat J(H,H)+2\hat J(H,G)+ 5\hat
J(G,G)\bigr]&(5.2)\cr}$$
(in fact this is the special case $N=4$ of the calculation reported in
ref.~[7]).
$$\eqalignno{V_{SF}&=3h^2\{ (2T-{1\over2}H)\hat I(T,T,H)
-{1\over2}G\hat I(T,T,G)+(T-G)\hat I(T,G,0)\cr &\quad+\hat J(T,T)-
\hat J(T,H)-2\hat J(T,G)\}, &(5.3)\cr}             $$
 $$\eqalignno{V_{SV}&={g^2\over{8\cos^2\th}}A(H,G,Z)+{g^2(1-2
 \sin^2\th)^2\over{8\cos^2 \th}}A(G,G,Z)+{1\over 2}e^2A(G,G,0)\cr&\quad
+{1\over4}g^2\bigl( A(H,G,W)+A(G,G,W)\bigr) -g^2\sin^4\th\/Z\/
B(Z,W,G)\cr&\quad-e^2\,W\/B(W,0,G)-{1\over2}g^2\,W\/B(W,W,H)-{g^2
\over{4\cos^2\th}}\/Z\/B(Z,Z,H)\cr&\quad+{g^2(1-2\sin^2\th)^2\over{
4\cos^2\th}}C(Z,G)+{g^2\over{8\cos^2\th}}\bigl(C(Z,H)+C(Z,G)  \bigr)
\cr&\quad+{1\over4}g^2\bigl(C(W,H)+3C(W,G)\bigr),&(5.4)\cr}           
$$
$$\eqalignno{V_V&=-{g^2\over4}\{            2\sin^2\th\/\Delta(W,W,0)+2\cos^2
\th\/\Delta(W,W,Z)-2\Sigma (W,W)\cr&\quad -4\cos^2\th\/\Sigma (W,Z)
+2A(0,0,W)+\cos^2\th\/A(0,0,Z)\}&(5.5)\cr}          $$
(The last two terms represent the ghost contribution).
$$\eqalignno{V_{FV}&=             -3\sum_f\bigl[(v^2_f+a^2_f)D(F,F,Z)+
(v_f^2-a^2_f)E(F,F,Z)\bigr]\cr&\quad-{3\over2}g^2(n_G-1)\/D(0,0,W)
-{1\over2}g^2 n_G D(0,0,W)-{3\over2}g^2\,D(T,0,W)\cr
&\quad -(4g_3^2+{4\over3}e^2)\bigl[
D(T,T,0)+E(T,T,0)\bigr].&(5.6)\cr}           $$
In Eq.~(5.6) the sum over $f$ is over all quarks and leptons, and
$v_f$ and $a_f$ denote the vector and axial couplings to the Z
boson. Thus, for example,
$$v_t={g(1-{8\over3}\sin^2\th)\over{4\cos \th}}\quad \hbox{and}\quad a_t=
{g\over{4\cos\th}}\eqno(5.7)$$
where $\th$ is the usual weak mixing angle.
$F=0$ except for the top quark, when $F=T$. $n_G$ is the number of
generations; of course since we neglect all Yukawa couplings except
$h$, our calculations are really only applicable for $n_G=3$.
The various functions introduced in Eqs.~(5.4)--(5.6) are
defined as follows:
$$\eqalignno{A(x,y,z)&={1\over z}\{-4a^2\hat I(x,y,z)+(x-y)^2\hat
I(x,y,0)+(y-x-z)\Jhat(x,z)\cr&\quad +(x-y-z)\Jhat(y,z)+z\Jhat(x,y)
            +2z(x+y-{1\over 3}z)\kappa^{-1}J(z)\},&(5.8)\cr}         $$
$$\eqalignno{B(x,y,z)&={1\over {4xy}}\{(10xy+z^2+x^2+y^2-2xz-2yz)
\hat I(x,y,z)\cr&\quad+(2zx-x^2-z^2)\hat I(x,z,0)+ (2zy-y^2-z^2)\hat
I(y,z,0)\cr&\quad+z^2\hat I(z,0,0)+(x+y-z)\Jhat(x,y)\cr&\quad -y
\Jhat(x,z)-
x\Jhat(y,z)+6xy\kappa^{-1}(J(x)+J(y))\cr&\quad-8xy\epsilon I(x,y,z)\},&(5.9)
\cr}      $$
$$C(x,y)=3\Jhat(x,y)-2\epsilon J(x,y)-2y\kappa^{-1}J(x),\eqno (5.10)$$
$$\eqalignno{D(x,y,z)&={1\over z}\{-(x^2+y^2-2z^2+xz+yz-2xy)
\hat I(x,y,z)+(x-y)^2
\hat I(x,y,0)\cr&\quad
-2z\Jhat(x,y)+(2z+y-x)\Jhat(x,z)\cr&\quad+(2z+x-y)\Jhat(y,z)
+{2\over 3}z(2z-3y-3x)\kappa^{-1}J(z)\cr&\quad +2\epsilon z\bigl[(x+y-z)
I(x,y,z)+J(x,y) -J(y,z)-J(x,z)\},&(5.11)\cr}$$
$$E(x,y,z)=\sqrt{xy}\big(6\hat I(x,y,z)+4\kappa^{-1}J(z)-4\epsilon I(x,y,z)),
\eqno(5.12)$$
$$\Delta(x,y,z) =\hat \Delta (x,y,z)
+\hat\Delta (y,z,x)+\hat\Delta (z,x,y)\eqno(5.13)
$$
where
$$\eqalignno{\hat\Delta(x,y,z)&={1\over {4xyz}}\{(x^4-8x^2yz-2x^2y^2+
32a^2xy)\hat I(x,y,z)\cr&\quad-((x^2-y^2)^2+8(x-y)^2xy)\hat I(x,y,0)
+x^4\hat I(x,0,0)\cr&\quad+z\Jhat(x,y)(9x^2+9y^2-9xz-9yz+13xy-z^2)-(
{4\over d}-1)xyzJ(x,y)\cr&\quad+8(x-y)^2xy\epsilon I(x,y,0)
+8z(xz+yz-x^2-y^2-xy)\epsilon J(x,y)\cr&\quad -4xyz({25\over3}x+6y+6z)\kappa^{-1}J(x)
-32a^2xy\epsilon I(x,y,z)\},&(5.14)\cr}         $$
and
$$\Sigma (x,y)={27\over 4}\Jhat(x,y)+\Bigl({(d-1)^3\over d}-{27\over
4}\Bigr)J(x,y)-{9\over 2}\kappa^{-1}(xJ(y)+yJ(x)).\eqno (5.15)$$
The $a^2$ that appears in Eq.~(5.8), Eq.~(5.14) is defined in Eq.~(4.12).
 The apparent singularities in, for example, $A(x,y,z)$ for $z=0$ are
easily seen to be spurious.
 
It is straightforward (but tedious) to substitute Eqs.~(5.8)--(5.15)
in Eqs.~(5.2)--(5.6). The resulting expression is not particularly
illuminating, however, and so we do not present it. Clearly
evaluation by computer will be necessary when we come to
applications, in any event.
\vskip 1in
\line{\bigbf 6. The renormalisation group.\hfil}
\line{\hfil}
$V(\ph)$ obeys the following RG equation:
$$\bigl(\mu{\pd\over{\pd\mu}}+\sum_i\beta_i{\pd\over{\pd\lambda_i}}-\gamma  \ph
{\pd\over{\pd\ph}}\bigr)V=0\eqno(6.1)$$
where $\lambda_i=\{m^2,\lambda,g,g',g_3,h\}$.
We begin by verifying Eq.~(6.1) at leading order. We have
$${\cal D}^{(1)}V^{(0)}=-\mu{\pd\over{\pd\mu}}V^{(1)}\eqno(6.2) $$
where
$${\cal D}^{(n)}=\sum_i\beta^{(n)}_i{\pd\over{\pd\lambda_i}}-\gamma^{(n)}  \ph {\pd
\over
{\pd\ph}}.$$
Using Eq.~(2.4), Eq.~(6.2) becomes:
$$\kappa{\cal D}^{(1)}V^{(0)}={1\over2}\bigl( H^2+3G^2-12T^2+6W^2+
 3Z^2\bigr).\eqno(6.3)$$
Note that there is a $\ph$-independent term on the RHS of Eq.~(6.3). We
therefore need to add a suitable term to $V^{(0)}$ of the form $f(\lambda_i)
m^4$. Equivalently, of course, one can simply redefine $V^{(1)}$ by
subtracting it at $\ph=0$[16].
This just shifts the potential by a constant, and has no consequences
for the considerations of this section. It will, however, contribute
non-trivially to the RG-improved form of the potential[17].
 
Comparing coefficients of $\ph^4$
                  and $\ph^2$, we obtain
$$\kappa (\beta^{(1)}_{\lambda}-4\lambda\gamma^{(1)})={1\over4}(16\lambda^2-144h^4+9g'^4+
18g^2g'^2+27g^4)\eqno(6.4)$$
and
$$\kappa (\beta^{(1)}_{m^2}-2m^2\gamma^{(1)})=2m^2\lambda.\eqno(6.5)$$
Substituting
$$\kappa \gamma^{(1)}={1\over4}(12h^2-9g^2-3g'^2)\eqno (6.6)$$
we obtain the well known results
$$\eqalignno{\kappa\beta ^{(1)}_{\lambda}&=4\lambda^2+12\lambda h^2-36h^4-9\lambda g^2
-3\lambda g'^2\cr
    &\quad+{9\over4}g'^4+{9\over2}g^2g'^2+{27\over4}g^4
                      &(6.7)\cr}       $$
and
$$\kappa\beta^{(1)}_{m^2}=m^2\bigl( 2\lambda+6h^2-{9\over2}g^2-{3\over2}g'^2
\bigr).\eqno (6.8)$$
 
At the two-loop level we have
$${\cal D}^{(2)}V^{(0)}=-\mu{\pd\over{\pd\mu}}V^{(2)}-{\cal D}^{(1)}V^{(1)}
\eqno(6.9)$$
The evaluation of $(\mu\pd/\pd\mu)V^{(2)}$ using the results of
section
5 is a straightforward application of the following results:
$$\eqalignno{\kappa ^2\mu{\pd\over{\pd\mu}}\hat I(x,y,z)&=2(L_1-2c),
&(6.10a)
\cr \kappa^2\mu{\pd\over{\pd\mu}}\Jhat(x,y)&=2xy(2-\overline {\ln}x
-\overline{\ln}y),&(6.10b)\cr  \kappa^2 \mu {\pd \over{\pd\mu}}\{\epsilon
I(x,y,z)\}&=
-2c,&(6.10c)\cr  \kappa^2\mu{\pd\over{\pd\mu}}\{\epsilon J(x,y)\}&=4xy,
&(6.10d)\cr
\kappa \mu {\pd\over{\pd\mu}}J(x)&=-2x.&(6.10e)\cr}$$
Using Eq.~(6.10) and Eqs.~(5.2)--(5.6) we find that
$$\eqalignno{\kappa^2 \mu {\pd\over{\pd\mu}}V^{(2)}&=\ph^4\{{19\over 18}
\lambda^3+2h^2\lambda^2-7h^4\lambda-12h^6+{8\over3}h^4g'^2+32g^2_3h^4+g'^4h^2\cr
   &\quad -3g^2g'^2h^2+{3\over32}\lambda g^4 +{1\over 32}\lambda g'^4
+{7\over16}\lambda g^2g'^2-{3\over2}\lambda^2g^2-{1\over2}\lambda^2g'^2
-{139\over{16}}g^6\cr
 &\quad +{5\over4}n_G\bigl( g^6+{1\over3}g^4g'^2+{5\over9}g^2g'^4+
 {5\over9}g'^6\bigr)+{23\over{48}}g^4g'^2+{143\over{96}}
g^2g'^4+{35\over96}g'^6\}\cr&\quad +m^2\ph^2\{
4\lambda^2+12\lambda h^2-18h^4
-{7\over2}\lambda(3g^2+g'^2)\cr&\quad    -{9\over 16}g^4+{15\over8}
g^2g'^2-{3\over{16}}g'^4\}\,\, +\hbox{logarithmic terms}.&(6.11)\cr}         $$
 
The logarithmic terms in Eq.~(6.11) are not given explicitly
because they simply {\it cancel} an identical set of terms
from ${\cal D}^{(1)}V^{(1)}$. This provides an excellent check on the
calculation, since there are terms of the form $\ln H$, $\ln G$,
$\ln T$, $\ln W$  and $\ln Z$ which must cancel separately.
 
The calculation of ${\cal D}^{(1)}V^{(1)}$ is straightforward; the
necessary one-loop RG functions are given by Eq.~(6.6)--(6.8)
and:
$$\eqalignno{\kappa \beta^{(1)}_h &= h\bigl( {9\over2}h^2-{17\over{12}} g'^2
-{9\over 4}g^2-8g_3^2\bigr),\cr
\kappa \beta_{g'} &={5\over3}g'^3\bigl({4\over 3}n_G+{1\over{10}}\bigr),
\quad
\kappa\beta_g=g^3\bigl({4\over3}n_G-{43\over6}\bigr).& (6.12)\cr}$$
We hence obtain from Eq.~(6.9) that
$$\eqalignno{\kappa^2\bigl(\beta^{(2)}_{\lambda}-4\lambda\gamma^{(2)}\bigr)
&= -{28\over 3}\lambda^3-24\lambda^2h^2+6\lambda^2(3g^2+g'^2)
+24\lambda h^4 +
{99\over4}\lambda g^4\cr &+{15\over 2}\lambda g^2g'^2
+{33\over{ 4}}\lambda g'^4+180h^6
-192h^4g^2_3 -16h^4g'^2 -{27\over2}h^2g^4\cr &+63h^2g^2g'^2
-{57\over 2}h^2g'^4 +3\{            \bigl(
{497\over8}-8n_G\bigr) g^6-\big({97\over{24}}+{8\over3}n_G\bigr)g^4g'^2
\cr &-\bigl({239\over24}+{40\over9}n_G\bigr)g^2g'^4-
\bigl({59\over{24}}+{40\over9}n_G\bigr) g'^6\}&(6.13)\cr}$$
and
$$\eqalignno{\kappa^2\bigl( \beta^{(2)}_{m^2}-2m^2\gamma^{(2)}\bigr)&=2m^2
\{             -\lambda^2-6\lambda h^2+2\lambda(3g^2+g'^2)\cr&\quad
+{63\over {16}}g^4+{3\over8}g^2g'^2+{21\over{16}}g'^4    \}.&(6.14)\cr}$$
 
We now require $\gamma^{(2)}$. For a general gauge theory (and in a
general covariant gauge) this is given in ref.~[18]; we have
merely to specialise to the Landau gauge and the SM, where care
must be taken to keep track of the factors. The result is:
$$\eqalignno{\kappa^2\gamma^{(2)}= &{1\over6}\lambda^2
-{27\over4}h^4+20g^2_3h^2+{45\over8}g^2h^2+{85\over{24}}g'^2h^2\cr
&+\bigl({5\over2}n_G-{511\over{32}}\bigr)g^4
+{9\over{16}}g^2g'^2+\bigl({25\over{18}}n_G+{31\over{96}}\bigr)g'^4.
&(6.15)\cr}$$
Substituting Eq.~(6.15) in Eq.~(6.13) and  (6.14) we obtain the
final results:
$$\eqalignno{\kappa^2\beta^{(2)}_{\lambda}
= &-{26\over 3}\lambda^3-24\lambda^2h^2
+6\lambda^2(3g^2+g'^2)-3\lambda h^4+80\lambda g^2_3 h^2\cr 
&\quad +{45\over2}\lambda g^2 h^2
+{85\over6}\lambda g'^2h^2       +
\bigl(10n_G-{313\over8}\bigr)\lambda g^4+{39\over4}\lambda g^2 g'^2\cr&
\quad+\bigl({50\over9}n_G +{229\over{24}}\bigr)\lambda g'^4+180h^6
-192h^4g^2_3-16h^4g'^2
-{27\over2}h^2g^4\cr &\quad+63h^2g^2g'^2-{57\over2}h^2g'^4
+3\Bigl(\bigl( {497\over8}-8n_G\bigr)g^6-\bigl({97\over{24}}+
{8\over3}n_G\bigr)g^4g'^2\cr&\quad
-\bigl({239\over{24}}+{40\over9}n_G\bigr)g^2g'^4
-\bigl({59\over{24}}+{40\over9}n_G\bigr)g'^6\Bigr).&(6.16)\cr}$$
and
$$\eqalignno{\kappa ^2\beta^{(2)}_{m^2}&=2m^2\{ -{5\over6}\lambda^2-
6\lambda h^2+2\lambda (3g^2+g'^2)-{27\over4}h^4+20g^2_3h^2
+{45\over8}g^2h^2\cr &\quad +{85\over24}g'^2h^2+\left({5\over2}n_G
-{385\over{32}}\right) g^4+{15\over{16}}g^2g'^2
+ \left({25\over 18}n_G+ {157\over{96}}\right)g'^4\}.&(6.17)\cr}  $$
We can compare our result for $\beta^{(2)}_{\lambda}$, Eq.~(6.16),
directly with Eq.~(B.8) of ref.~[6]. Agreement  is complete
apart from
 
(i) the sign of the $\lambda g^2g'^2$, $\lambda g'^4$ terms.
 
(ii) the magnitude of the $\lambda g'^4$ term.
 
From the result of a \it general \/\rm gauge theory (Eq.~(4.3) of
ref.~[6], or see also ref.~[19]) it appears that these discrepancies
arise from errors in
 the reduction to the SM case in ref.~[6] rather than
in the general result.\footnote{\dag}{See note added.}
 Unfortunately the incorrect form
of Eq.~(B.8) has been applied by other authors to running
coupling analyses, e.g. ref.~[20]. The numerical effect of the error on
these calculations is probably small, however.
 The result for $\beta^{(2)}_{m^2}$ can be
verified from the general formulae of ref.~[19] and will be needed when
we consider the RG improved form of $V$.
 
At this point it is appropriate to comment on the work of
Alhendi[5]. He essentially reverses the procedure adopted in
this section, in order to deduce the $\mu$-dependent terms in
$V^{(2)}$ given the $MS$ $\beta$-functions. Unfortunately
his expressions for $\gamma^{(2)}$ and $\beta^{(2)}_{m^2}$ are
incorrect; he assumes for example (in our notation) that the
coefficient of the $g^4$ term in $\gamma^{(2)}$ is just ${1\over4}$
 of the coefficient of the $\lambda g^4$ term in $\beta^{(2)}_{\lambda}$.
This amounts to the assertion that there are no 1PI contributions
of this kind to $\beta^{(2)}_{\lambda}$, which is not correct (even in
the Landau gauge). Aside from this comparison would still be
difficult since he does not use the $MS$ form of $V^{(1)}$.
These problems aside, it is clear that by this method one can
derive the form of the potential in the region $\ph^2\gg \vert m^2\vert$
only.
\vskip 1in
\line{\bigbf 7. Conclusions and outlook\hfil}
\line{\hfil}
We have presented a calculation of $V^{(2)}(\ph)$ in the standard
model, using dimensional regularisation and minimal subtraction.
Applying the renormalisation group to the result led to numerous
checks and also expressions for $\beta^{(2)}_{\lambda}$ 
and $\beta^{(2)}_{m^2}$.We used differential equations and 
the method of characteristics
to find the relevant Feynman integral; an approach which seems to us
preferable to more traditional techniques.
 
Studies of the stability of the electroweak vacuum reviewed in ref.~[2]
suggest a limit of around $m_t\leq95GeV+0.6m_H$ for the top and Higgs
masses. Our calculation will enable us to probe further the sensitivity
of this result to radiative corrections. We will report these
calculations elsewhere; the following argument, however, suggests that
it will not be affected dramatically: Instead of solving the RG
equations to produce an ``improved'' $V$, suppose we take the
unimproved $V$ and simply set $\mu=\ph$. All couplings (and $m^2$)
then become functions of $\ph$; but this choice of $\mu$ means that
when $\ph^2\gg\vert m^2\vert $,
   the radiative corrections to $V$, although non-zero,
have no large logarithms. As long as the running couplings remain
perturbative, we may therefore expect the radiative corrections to cause
only a small change in the above limit. Perturbative confidence
in the limit means that
should it turn out that $m_t$ and $m_H$ fail to satisfy it, this would
provide compelling evidence for physics beyond the standard model.
\bigskip

\line{\it Note Added\hfil}
Our thanks to Mike
Vaughn for correspondence, and for confirming that 
the authors of ref.~[6] now agree
with our result for $\beta_{\lambda}^{(2)}$, Eq.~(6.16).   

The published version of this paper (and unfortunately also 
the published erratum) contain a number of typographical 
errors and omissions, which have been corrected here. 
These errors affected the result for $V^{(2)}$ (Eq.~(5.1)) 
and the result for $\beta^{(2)}_{m^2}$ (Eq.~(6.17)), but 
not the result for $\beta^{(2)}_{\lambda}$ (Eq.~(6.16)).

We are grateful to Peter Arnold, Steve Martin and Mingxing Luo who drew
most of these errors  to our attention.  See ref.~[21]-[23]; ref.~[22]
includes a generalisation of  the two loop effective potential
calculation to the case of  an arbitrary gauge theory.  \bigskip 
 
\line{\bigbf Acknowledgements\hfil}
\line{\hfil}
While part of this work was done, one of us (TJ) enjoyed the hospitality
of the Institute of Theoretical Physics, University of California at
Santa Barbara. We thank Marty Einhorn for conversations. This
research was supported in part by the National Science Foundation
under grant no. PHY89-04035, by a NATO collaboration research
grant, and by the SERC. Thanks also to Peter Arnold, Steve Martin and 
Mingxing Luo  for correspondence (as explained above).

\vskip 1in
\line {\bigbf References \hfil}
\line {\hfil}
\item {1. } S. Coleman and E. Weinberg, Phys. Rev. {\bf D7} (1973) 1888.
\item {2. } M. Sher, Phys. Rep. {\bf 179} (1989) 274.
\item {3. } M. J. Duncan, R. Phillippe and M. Sher, Phys. Lett. {\bf
B153} (1985) 165;
\item {   } M. Sher and H. W. Zaglauer, Phys. Lett. {\bf B206} (1988) 527;
\item {   } M. Lindner, M. Sher and H. W. Zaglauer, Phys. Lett. {\bf B228
} (1989) 139;
\item {   } J. Ellis, A. Linde and M. Sher, Phys. Lett. {\bf B252} (1990)
203.
\item {4. } K. Yamagishi, Phys. Rev. {\bf D23} (1981) 1880; Nucl Phys.
{\bf B216} (1983) 508;
\item {   } M. B. Einhorn and D. R. T. Jones, Nucl. Phys. {\bf B211}
(1983) 29.
\item {5. } H. Alhendi, Phys. Rev. {\bf D37} (1988) 3749.
\item {6. } M. E. Machacek and M. T. Vaughn, Nucl. Phys. {\bf B249}
(1985) 70.
\item {7. } C. Ford and D. R. T. Jones, Phys. Lett. {\bf B274} (1992)
409.
\item {8. } D. R. T. Jones, Nucl. Phys. {\bf B277} (1986) 153.
\item {9. } K. T. Mahanthappa and M. Sher, Phys. Rev. {\bf D22} (1980)
1711.
\item {10.} A. V. Kotikov, Phys. Lett. {\bf B254} (1991) 158; Phys. Lett.
{\bf B259} (1991) 314.
\item {11.} R. Jackiw, Phys. Rev. {\bf D9} (1974) 1686.
\item {12.} N. K. Nielsen, Nucl. Phys. {\bf B101} (1975) 173;
\item {   } R. Fukuda and T. Kugo, Phys. Rev. {\bf D13} (1976) 3469.
\item {13.} E. Braaten and J. P. Leveille, Phys. Rev. {\bf D24} (1981) 13.
\item {14.} K. G. Chetyrkin, A. L. Kataev and F. V. Tkachov, Nucl. Phys.
{\bf B174} (1980) 345;
\item {   } F. V. Tkachov, Phys. Lett. {\bf B100} (1981) 65;
\item {   } K. G. Chetyrkin and F. V. Tkachov, Nucl. Phys. {\bf B192}
(1981) 159;
\item {   } D. I. Kazakov, Phys. Lett. {\bf B133} (1983) 406; Teor. Math.
Fiz. {\bf 58} (1984) 210; {\it ibid} {\bf 62} (1985) 127;
\item {   } D. T. Barfoot and D. J. Broadhurst, Z. Phys. {\bf C41}
(1988) 81;
\item {   } D. J. Broadhurst, Z. Phys. {\bf C47} (1990) 115;
\item {   } N. Gray, D. J. Broadhurst, W. Grafe and K. Schilcher,
 Z. Phys. {\bf C48} (1990) 673.
\item {15.} I. S. Gradshteyn and I. M. Ryzhik, ``Tables of Integrals,
Series and Products'' (Academic Press, 1980).
\item {16.} M. B. Einhorn and D. R. T. Jones, in ref.~[4].
\item {17.} B. Kastening, Phys. Lett. {\bf B283} (1992) 287.
\item {18.} M. E. Machacek and M. T. Vaughn, Nucl. Phys. {\bf B222}
 (1983) 83.
\item {19.} I. Jack and H. Osborn, Nucl. Phys. {\bf B249} (1985) 472.
\item {20.} K. S. Babu and E. Ma, Z. Phys. {\bf C31} (1986) 45.
\item {21.} P. Arnold and L. G. Yaffe, Phys. Rev. {\bf D55} (1997) 7760.
\item {22.} S. P. Martin,  Phys.Rev. {\bf D65} (2002) 116003.
\item {23.} Mingxing Luo and Yong Xiao, in preparation.

\end